\documentclass[12pt,preprint]{aastex}
\begin{document}

\title{Photometry of the Stingray Nebula (V839 Ara) from 1889-2015 Across the Ionization of Its Planetary Nebula}
\author{Bradley E. Schaefer \& Zachary I. Edwards\affil{Physics and Astronomy, Louisiana State University, Baton Rouge, LA 70803}}

\begin{abstract}

Up until around 1980, the Stingray was an ordinary B1 post-AGB star, but then it suddenly sprouted bright emission lines like in a planetary nebula (PN), and soon after this the {\it Hubble Space Telescope} ({\it HST}) discovered a small PN around the star, so apparently we have caught a star in the act of ionizing a PN.  We report here on a well-sampled light curve from 1889 to 2015, with unique coverage of the prior century plus the entire duration of the PN formation plus three decades of its aftermath.  Surprisingly, the star anticipated the 1980's ionization event by declining from B=10.30 in 1889 to B=10.76 in 1980.  Starting in 1980, the central star faded fast, at a rate of 0.20 mag/year, reaching B=14.64 in 1996.  This fast fading is apparently caused by the central star shrinking in size.  From 1994-2015, the V-band light curve is almost entirely from the flux of two bright [OIII] emission lines from the unresolved nebula, and it shows a consistent decline at a rate of 0.090 mag/year.  This steady fading (also seen in the radio and infrared) has a time scale equal to that expected for ordinary recombination within the nebula, immediately after a short-duration ionizing event in the 1980s.  We are providing the first direct measure of the rapidly changing luminosity of the central star on both sides of a presumed thermal pulse in 1980, with this providing a strong and critical set of constraints, and these are found to sharply disagree with theoretical models of PN evolution.

\end{abstract}

\section{Background}

	The `Stingray Nebula' (V839 Ara, SAO 244567, CD -59$\degr$ 6479, CPD -59$\degr$ 6926, Hen 3-1357, and PN G331-12.1) is a unique case where an ordinary post asymptotic giant branch (post-AGB) star suddenly changed its appearance to that of a young planetary nebula (PN).  This ejection of the PN shell was roughly one millennium ago, but the nebula turned on suddenly around 1980 with some sharp increase in ionizing radiation.  The Stingray represents our one opportunity to actually watch the turn-on of a PN.
	
	Before 1980, four spectra of the Stingray have been published, all showing either very weak Balmer emission lines or no emission lines (see the timeline in Table 1).  During this time, as best seen in the 1971 spectrum (Parthasarathy et al. 1995), the spectrum was that of a normal B0 or B1 star, placed into luminosity class I or II, with possible weak emission at $H\beta$ only.  The star slowly began to attract attention, first for having just some $H\alpha$ emission (Henize 1976), then as an {\it IRAS} far-infrared source selected out as a proto-planetary nebula (Volk \& Kwok 1989; Parthasarathy \& Pottasch 1989), then as a star that had a sudden appearance of a PN shell within the previous few years (Parthasarathy et al. 1993).  By 1990 and 1992, the Stingray optical spectrum was dominated by very bright and narrow [OIII] emission lines, plus other lines that are characteristic of a young PN (Parthasarathy et al. 1993; 1995).  The stark difference between the 1971 spectrum (a B1I star with weak emission only) and the 1990 spectrum (very bright PN emission lines) impresses that the Stingray is evolving fast and apparently the PN has just turned-on.
	
	From the start of the 1980's event with the fast turn-on of the PN spectrum, the Stingray has been fast evolving (see Table 1).  At the start, a B0 spectrum gives a surface temperature of around 30,000 K.  From the time of the first {\it International Ultraviolet Explorer} ({\it IUE}) spectrum, the effective temperature of the star has been heating up greatly (Parthasarathy et al. 1995), going from 38,000 K in 1988 to 55,000 K in 2006 (Reindl et al. 2014).  The surface gravity has increased from $10^{4.8}$ cm s$^{-2}$ in 1988 to $10^{6.0}$ cm s$^{-2}$ in 2006 (Reindl et al. 2014).  The mass ejection rate has decreased from $10^{-9.0}$ $M_{\odot}$ year$^{-1}$ in 1988 to $10^{-11.6}$ $M_{\odot}$ year$^{-1}$ in 2006 (Reindl et al. 2014).  The ultraviolet continuum brightness has fallen by a factor of three from 1988 to 1994 (Parthasarathy et al. 1995; Feibelman 1995), continued to fall by another factor of three to 1997, but then brightened by a factor of two in 2002 and 2006 (Reindl et al. 2014).  We have a clear picture of a star in 1988 with a high stellar wind, that rapidly tapered off, while the star shrunk in size (by a factor of four) and and heated up.  The shrinking and heating of the central star is what would be expected for a simple picture of a stellar core following along the evolutionary track that leads to a white dwarf.
	
	The first distance measure to the Stingray was 5.64 kpc (Kozok 1985b), based on a presumed absolute magnitude appropriate for the star's color, and as such this estimate has a very large real uncertainty.  Fresneau et al. (2007) measure the proper motion of the Stingray, and from this derive a statistical parallax of 1.21$\pm$0.21 milli-arc-seconds (for a distance of 830 pc and an uncertainty of 17\%), while the possible deviation from their statistical model makes for a possibly large error in this distance.  The best distance to the Stingray is based on the luminosity calculated from the measured temperature and surface gravity, with resulting distances of 1.6$^{+0.8}_{-1.2}$ kpc (Reindl et al. 2014) and $\approx$1.8 kpc (Arkhipova et al. 2013).  The interstellar extinction has $E(B-V)$ equal to 0.20$\pm$0.05 mag (as based on the 2200\AA~feature in the ultraviolet) and 0.14 mag (based on the observed Balmer decrement) (Parthasarathy et al. 1993).  The measured extinction has not changed significantly from 1980 to 2011 (Reindl et al. 2014).  The deduced mass for the star is variously given as $<$0.55 $M_{\odot}$ (Reindl et al. 2014) or 0.2+0.59 $M_{\odot}$ for the ionized mass plus the core mass (Bobrowsky 1994).

	The expanding PN shell was resolved in 1992 by the {\it Hubble Space Telescope} ({\it HST}) to be bipolar shaped with an embedded `equatorial ring', where the largest radius was 0.8 arc-seconds (Bobrowsky 1994; Bobrowsky et al. 1998).  The inclination of the ring is 56$\degr$.  Just inside this ring, a V=17.0$\pm$0.2 mag star is a possible wide companion star, making the Stingray a binary with a separation of $\sim$2200 AU.  The presence of the companion star has been suggested to have some effect on the formation of the bipolar shape, but the very wide separation makes it hard for any such effects to be substantial.  Importantly, the visible size of the PN requires an ejection many hundreds or thousands of years ago.  Given an angular size of 1.15", a distance of 1.6 kpc, and an expansion velocity of 8.4 km s$^{-1}$ from the [OIII] line width, we get a kinematic age of the main PN shell to be 1013$^{+488}_{-793}$ years (Reindl et al. 2014).  So we have a clear picture that the PN shell was ejected about a millennium ago, but was only ionized in 1980.

	The Stingray has reasonable coverage from around 1920 to the present with ground-based spectroscopic monitoring, from 1988 to 1996 with {\it IUE} spectra, and from 1992 to 2000 with {\it HST} ultraviolet spectroscopy and high-resolution imaging.  But no one has reported any photometry past a few isolated magnitudes, all with large problems, spread over various magnitude systems.  We realized that a full light curve was needed, and we had means for measuring full light curves all the way back to 1889, so as to cover over eight decades of time before the 1980's ionization event, the transition time of the PN turn-on, and three decades of the coasting phase after the turn-on.  Nor has anyone looked for an outer shell far away from the central star (such as are seen around many ordinary PNe), so we wondered whether the Stingray has any far outer shells.  This paper reports on our results of the light curve of the Stingray from 1889 to 2015, plus our searches for any outer halo. 

\section{Photometry}

	To get broad-band magnitudes for the Stingray, we have pulled from a wide variety of sources; the Harvard photographic plate collection from 1889-1989, the visual magnitude estimates of Albert Jones as archived by the {\it American Association of Variable Star Observers} ({\it AAVSO}) from 1994 to 2007, the {\it All Sky Automated Survey} ({\it ASAS}) from 2001 to 2009, the {\it AAVSO} telescopes going into the {\it AAVSO Photometric All-Sky Survey} ({\it APASS}) from 2011 to 2015, plus our own photometry from CCD images with DECam on the Cerro Tololo 4-m Blanco telescope from 2014.  In all, we have 1026 magnitudes, mainly in B and V, with good coverage from 1889 to 2015.  We have added 15 magnitudes from the literature or derived by us from the literature, all on 6 nights from 1969 to 1996.  All 1041 magnitudes are presented in Table 2.
	
	A critical realization is that the magnitudes {\it before} 1980 are all measures of the continuum of the star, while all magnitudes {\it after} roughly 1989 are essentially measures of the [OIII] emission line strengths.  The reason is that before this cutoff date there were no significant emission lines seen in the spectrum so the brightness is of the observed stellar continuum, whereas after that date we see the optical fluxes dominated by emission lines.  So before 1980, we are measuring the brightness of the star alone, and this depends only on the effective size and temperature of the photosphere minus any extinction by dust.  After 1989, the emission lines dominate, with the brightness depending on both emission line fluxes and the exact spectral sensitivity of the detector.  The nebular light in the V-band is entirely dominated by just three emission lines; [OIII] at 5007\AA~as the brightest with 77\% of the V-band detected flux, [OIII] 4959\AA~with 21\% of the flux, and H$\beta$ at 4861\AA~as the faintest with only 2\% of the flux.  So Jones, {\it ASAS}, and {\it APASS} are essentially reporting the brightness of the [OIII] emission lines. 
	
	A further realization from this is that there will likely be systematic differences in the post-1980's magnitudes from observer to observer.  For example, the exact width of the spectral sensitivity curve for Albert Jones' eyes is somewhat different from the {\it ASAS} CCD/telescope/atmosphere combination, so Jones will detect somewhat more [OIII] light relative to what he detects for the nearby comparison stars as compared to the {\it ASAS} measures.  These systematic observer-to-observer offset are real, and apparently at the quarter-of-a-magnitude level.

	Another critical realization is that the emission line spectrum for V839 Ara makes its measured brightness respond slightly differently to atmospheric extinction than do the brightnesses of the nearby comparison stars.  The reason is that V839 Ara has all of its light on the blue-side of the V-band, while the ordinary comparison stars have a continuum that fairly evenly covers the entire V bandpass.  Ordinary atmospheric extinction dims the blue side of the V bandpass more than the red side.  Extinction from the ground changes on many timescales due to differing airmass (as the Stingray is looked at with varying zenith distances) and with ordinary changes in the atmosphere aerosol content.  Thus, if atmospheric extinction increases for any reason, then the blueish emission light of V839 Ara will be dimmed somewhat more than the light from the normal comparison stars will be dimmed.  All of the ground-based photometry reported here is straight differential photometry with respect to an ensemble of nearby comparison stars, so V839 Ara will appear to slightly dim in our derived V magnitude as the atmospheric extinction increases.  This effect cannot be taken out with the usual color terms derived from the comparison stars for the CCD/filter/telescope combination, and indeed, an exact correction is not possible without more information on the atmosphere and spectral sensitivities than is available.  So we are left with a relatively small atmospheric source of variability superposed on the real underlying light curve.  From the {\it APASS} time series, we see that the effect can get up to around 0.05 mag in amplitude.  For all other ground-based telescopes, we cannot recover from this atmospheric variation, so we just have to acknowledge that the measured magnitudes have a non-astrophysical variability of roughly 0.05 mag superposed.  This problem arises from the nearly-unique nature of the Stingray's emission line spectrum.  Importantly, this mechanism does {\it not} apply to the pre-1980s Harvard magnitudes, because the star dominated with no significant emission lines. Fortunately, for the post-1980s light curve, fast variations are impossible (because the visible light is coming from emission lines spread out over a region one light-month in size), so we can use our many magnitudes to get time-averaged variations with atmospheric effects well-averaged out.
	
\subsection{Harvard Plates}

	The Harvard College Observatory has a collection of over 500,000 archival photographic plates that cover the entire sky from 1889 to 1953, plus a smaller collection from around 1969 to 1989.  (The years 1953-1969 are the notorious Menzel Gap.)  These plates are mostly in the B band, with limiting magnitudes typically ranging from 14 to 18 mag.  Any one position will have typically from 1000 to 4000 plates for coverage.  The {\it Digital Access to a Sky Century @ Harvard} program ({\it DASCH}) is currently $\sim$10\% through digitizing all the Harvard plates, but it will reach the Stingray (in the constellation Ara at low galactic latitude) only some years from now.
	
	We have constructed a light curve from 108 Harvard patrol plates, all in the B band.  The plates were all from the B series as well as on patrol plates of the RB, AM, AX, and DSB series.  (We could have measured many more magnitudes for such a bright star as the Stingray, but the patrol plates taken have the complete coverage in time and already well-define the star's variability, so adding more plates would provide little new information.)  The comparison stars were chosen from nearby stars of similar magnitude, with their B magnitudes taken from the {\it APASS} survey of the {\it AAVSO}.  The Harvard plates formed part of the original definition of the B magnitude system, and they have been measured many times to have a near-zero color term for transformations from their native system to the Johnson B magnitude system.  With this, we can be very confident that our resultant magnitudes are in the Johnson B magnitude system.
	
	The visual comparison of targets to the comparison stars is a long well-developed practice, now largely lost amongst living astronomers.  Nevertheless, extensive experimentation over the last thirty years has proven that the visual estimation of magnitudes has an accuracy essentially equal to that obtained from digital scans and from iris diaphragm photometers.  Importantly, an experienced eye will produce a real uncertainty that is more than a factor from 1-times to 3-times better than that produced by the {\it DASCH} photometric pipeline (e.g., Schaefer 2014a, 2014b).  Given our very long and deep experience at visual estimation, plus its great speed, simplicity, and low-cost, the visual estimation method is to be preferred.  For an average case with a good sequence and a target well above the plate limit (as for the case of the Stingray), the real one-sigma error bar is close to $\pm$0.10 mag.
	
	The Harvard plates light curve from 1889 to 1989 is presented in Table 2 and Figure 1.  The behavior of the Stingray was startlingly unexpected.  From 1889 to 1980, the star slowly faded, with significant modulation.  The fading is consistent with a linear decline in magnitude, going from B=10.30 in 1889 to B=10.76 in 1980, for a decline at a rate of 0.0051 mag/year.  Superposed on this linear decline are apparent fluctuations on the decadal timescale with a total amplitude of near half a magnitude.  Then, starting suddenly around 1980, the Stingray central star faded fast at least until 1989.  The rate of decline was approximately 0.20 mag/year.
	
\subsection{Albert Jones}

	From early 1994 until the middle of 2007, Albert Jones has reported 128 visual measures of the Stingray.  These are all visual measures, through his 12.5-inch reflector telescope in Nelson New Zealand, made using a sequence of comparison stars that is accurate on the modern magnitude scale.  Thus, his measures are closely on the Johnson V-magnitude system.  Jones' magnitudes were reported through the Variable Stars Section of the {\it Royal Astronomical Society of New Zealand}, and are now available in the {\it AAVSO} database.
	
	Albert Jones has been the world's best observer of variable stars for much of the last sixty years.  (He died in 2013.)  Some of his exploits include the discovery of the highly-important SN 1987A, the discovery of two recurrent novae eruptions (T Pyx in 1966 and V3890 Sgr in 1990), and the discovery of two comets.  In the early 1990's, Janet Mattei (then {\it AAVSO} Director) pointed him out as being the all-time best variable star observer with incredibly accurate eyes, while Daniel W. E. Green (now the Director of the {\it Central Bureau for Astronomical Telegrams}) pointed him out as being the world's best observer of comet magnitudes.  And he not only has an incredible accuracy, but he also has made over 500,000 variable star magnitude measures, nearly a factor of two times more observations than any other visual observer.  The authors of this paper have used Jones' observations in many prior studies for a wide range of stars, and we have always been deeply impressed by both his quality and quantity of observations, as well as by his characteristic of always looking at the fun, exciting, and useful variables.  Results from our prior work, and from others, give Jones as having a one-sigma photometric accuracy of 0.05 to 0.10 mag.  The reason for this recital is to point to Jones' reliability and accuracy.
	
	The RMS scatter of the differences between successive magnitudes in the light curve is 0.21 mag for pairs within 10 days.  Jones is only measuring the [OIII] emission lines, and these cannot vary substantially on such fast time scales.  (This is because the [OIII] light comes from all around the planetary nebula, with near-zero from the central star, and the nebula is about a light-month in radius.)  As 0.21 mag is the difference between two magnitudes, the one-sigma uncertainty for measuring one magnitude will be 0.15 mag.  This variation is significantly larger than Jones' photometric accuracy.  So we take the excess variance to be due to the effects of variations in the atmospheric extinction, where the emission lines from V839 Ara are dimmed more than the continuum light from the comparison stars.  Thus, the extinction effect is somewhat smaller than 0.15 mag in size.
	
	Jones' visual light curve is plotted in Figure 2, with circles representing his measures.  His light curves shows a steady decline with superposed apparently-random fluctuations.  His observations are reasonably fit with a linear decline, from V=10.71 in 1994 to V=11.77 in 2007, for a decline rate of 0.081 mag/year.  The RMS scatter around this linear decline is 0.22 mag.  
	
\subsection{ASAS}
	
	{\it ASAS} has been running an all-sky survey, measuring V magnitudes for approximately 10 million stars on a nightly basis (Pojmanski 2002).  The telescopes all have lenses of aperture 200 millimeters and f-ratio of 2.8, forming an 8.5$\degr$$\times$8.5$\degr$ field of view, while the CCD has a pixel size of 15".  With 180 second integrations, the limiting magnitude is roughly V=14.  For observations in V of the Stingray, the telescope is at Las Campanas Observatory in Chile.  The publicly available light curve is from early 2001 to late 2009, and we will have 425 V magnitudes.
	
	For the reported {\it ASAS} V magnitudes, we selected out only the grade A and B values.  These magnitudes show a slow secular decline from 2001 to 2009, plus a large scatter superposed on top.  Uncrowded neighboring stars of similar brightness show a good flat light curve with an RMS scatter of 0.06 mag.  The publicly reported magnitudes are divided into ten blocks, where each block was made with a different field center and slightly different positions for the centers of the photometry apertures.  Each field center produces a scattered light curve that parallels the light curves from the other centers, yet the field centers have offsets that vary by 1.6 mag (with an RMS scatter in the intercepts of 0.57 mag).  The cause for this variations is that a V=10.73 star is 35" from the Stingray, so small changes in the center of the photometry aperture will make for a varying contribution from the nearby star.  To pull out this effect, we have constructed a model with a presumed Gaussian point spread function (PSF) for the {\it ASAS} star profiles.  For the known aperture radii, the known star positions, and the known centers for the photometry aperture, this model has only one fit parameter, the Gaussian width of the PSF.  The {\it ASAS} literature gives typical values from 10" to 15" (e.g., Pojmanski 2002).  We have made a chi-square fit for the reported brightness for each of the ten field centers and the five aperture diameters (30", 45", 60", 75", and 90").  Our best fit is with a Gaussian sigma value of 14".  With this value plus our model, we have derived the V magnitude for the Stingray.
	
	The resultant light curve still shows substantial scatter.  For pairs of magnitudes taken within 10 days of each other (so that the nebular [OIII] light cannot vary significantly), the RMS scatter of the magnitude differences is 0.31 mag.  The one-sigma measurement uncertainty for one magnitude is 0.22 mag.  This large scatter is much larger than dictated by photon statistics.  The cause for this scatter can be due to the atmospheric extinction variations (as discussed previously) as well as the circumstance of having imperfect correction for a relatively bright star just near the edge of the photometry aperture.  Small variations in the PSF width and the center of the photometry aperture can make for substantial changes in the derived brightness for the Stingray, and we have no way to recover this from the publicly available data.
	
	The {\it ASAS} light curve from 2001 to 2009 (see Figure 2) shows a steady decline, with this being highly significant despite the substantial measurement errors.  This decline is consistent with a linear change in the magnitude, going from 11.41 in 2001.0 to 12.53 in 2010.0, at a rate of 0.124 mag/year.  The RMS scatter around this trend line is 0.23 mag.
	
	Figure 2 shows us that the ASAS light curve has a similar slope from 2001-2009 as does Jones from 1994-2007, but that there is an offset by 0.2 to 0.4 mag in the overlap time interval.  It is possible that our model of the Stingray produced a systematic underestimate of its contribution to the smallest photometry aperture.  However, we think that it is more likely due to Jones and ASAS having different spectral sensitivities.  That is, the photochemicals in Jones' eye will produce a distinctly different relative sensitivity for the [OIII] lines (when compared to the mean sensitivity for the continuum from the comparison stars) than will the ASAS CCD-plus-filter.  In this case, the observed offset is just the result of normal color terms between two different detectors, with this being exaggerated for an emission line source.  In all cases, the photometric uncertainties are below the quarter-magnitude level, and we can easily see the overall behavior of the Stingray.
		
\subsection{APASS}

	{\it APASS} is measuring B, V, g', r', and i' magnitudes of all stars in the sky with approximately $10<V<16$, with several telescopes in both the northern and southern hemisphere.  This has provided us with a reliable source of comparison stars for our differential photometry.  This has also provided us with magnitudes in the five filters on two nights in 2011 (see Table 2).
	
	At our request, A. Henden has put the Stingray in the queue for time series photometry on the 0.61-meter Optical Craftsmen Telescope at the Mount John Observatory in New Zealand.  The 1-minute CCD integrations were through a Johnson V filter on the nights of 23, 26, and 27 March 2015.  The quoted magnitudes were based on differential aperture photometry with respect to 5-8 nearby comparison stars previously calibrated with {\it APASS}.  On the first night, as the Stingray rose from an altitude of 31$\degr$ to 54$\degr$, the apparent brightness of the Stingray suffered a dip in brightness by about 0.05 mag for one hour.  The Stingray was constant, with an RMS scatter of 0.008 mag, on the other two nights.  During the time of the dip, the statistical error for the star increased from 0.012 mag up to 0.06 mag, indicating that some large atmospheric extinction had dimmed the target by much more than 0.05 mag.  For the time through the dip and for all the nights, we found a tight correlation between the calculated statistical error and the differential magnitude of the target.  This has the easy interpretation that ordinary changes in the atmospheric aerosols substantially dimmed both the Stingray and its comparison stars (making for large statistical errors), while the Stingray's [OIII] emission lines (all on the blue side of the V bandpass) were dimmed more than the continuum light from the comparison stars (making for the small dip in the differential magnitude).  This effect is undoubtedly happening on all time scales to all ground-based observers, but we can see this effect only in a time series from one telescope.  With this correlation, we can reduce all the {\it APASS} magnitudes to those of some constant condition with minimal aerosols.  For the measures with statistical error greater than 0.020 mag, this correction could not be done with high accuracy, so we have chosen to delete these data.  The result is a time series on three nights with 352 V-band magnitudes (see Table 2), with nightly averages of 12.62, 12.57, and 12.61.
	
	The {\it APASS} magnitudes extends the V-band light curve past 2009 up until March 2015 (see Figure 2).  From the {\it APASS} magnitudes alone, we see a significant decline from 2011 to 2015.  This is just a smooth extension of the earlier decline.  The various observers are not expected to report the same V magnitudes to within a quarter of a magnitude or so, due to the emission line nature of the Stingray.  Nevertheless, all observers show a consistent and highly significant steady fading from 1994-2015.  This fading goes from V=10.71 in 1994.1 to V=12.61 in 2015.3, for an average decline rate of 0.090 mag/year.
	 		
\subsection{DECam on Cerro Tololo 4-m}

	With the Cerro Tololo 4-meter Blanco Telescope, DECam CCD images were taking on 30 June 2014, the main goal being to search for any outlying faint optical shell. The large size (9'x17') of one individual chip allows for V839 Ara, comparison stars, and for any far out shell to be contained on a single CCD (out of the 64 CCD array).  Observations were made using the the u', g', r', i' (which are the Sloan Digital Sky Survey, SDSS, filters) with pairs of exposure time of 20 and 90 seconds in each band.
	
	Standard procedures were used in the IRAF data reduction package to extract magnitudes for V839 Ara utilizing the PHOT package.  Differential photometry was performed on V839 Ara using multiple {\it APASS} comparison stars, carefully avoiding saturated stars, and only with the short exposures.  These comparison stars were chosen such that they fell on the same chip as V839 Ara, have similar brightness, and not crowded by surrounding stars. Since {\it APASS} only provides information for g', r', \& i' filters, a magnitude for the u' observations could not be derived.  We measured the magnitudes of V839 Ara to be 12.52$\pm$0.01, 12.18$\pm$0.01, \& 13.35$\pm$0.01 for g', r', \& i' respectively.  The formal statistical error bars are smaller than 0.01 mag, while various systematic sources of error are likely around the 0.05 mag level.
		
\subsection{Additional Optical Magnitudes}

	Hill et al. (1974) and Kozok (1985a) report on UBV photometry on four separate nights.  These magnitudes are of the central star, with any contributions from emission lines and the PN being negligibly small.  These four B magnitudes closely fit with the light curve from the Harvard plates.
	
	Bobrowsky et al. (1998) used {\it HST} to resolve the central star and report V=15.4 as deduced from the flux measured in a continuum filter centered at 6193\AA.  This magnitude is of the central star alone (with no emission lines and no PN shell) was made in March 1996.  Around this time, Albert Jones was reporting the Stingray to be V=10.8 mag.  The difference of 4.6 mag is because Jones is including the shell light, almost entirely [OIII] emission lines.  This shows that the two [OIII] lines are contributing $\sim$70$\times$ as much flux as the star over the entire broad V-band.
	
	Reindl et al. (2014) report on an {\it HST} spectrum from March 1996 with the {\it FOS}.  We have taken the continuum flux for the center of the B- and V-bands, and converted these into magnitudes.  We get B=14.64 and V=14.96, with this applying to the star alone.  The V magnitude is 0.44 brighter with the {\it FOS} spectrum than with Bobrowsky's continuum flux, despite there being only two days of separation in time.  It is unclear if the difference arises from variability or from the ordinary uncertainties in extracting magnitudes with two different non-standard methods.
	
	Other published magnitudes (like the estimates from the {\it Cape Durchmusterungen} and {\it HD} catalogs) have problems with their comparison stars that can be as large as one magnitude, and hence cannot be used with any useful accuracy or confidence.  The presentation of the {\it ASAS} magnitudes by Arkhipova et al. (2013) has not realized any of the problems or solutions caused by the varying centers and the nearby star, so their light curve is now to be replaced by our light curve described in Section 2.3.
		
\subsection{Infrared and Radio Brightnesses}
	
	From the 1992 Ph.D. thesis of P. Garcia-Lario (as quoted in Parthasarathy et al. 1993) the J, H, and K magnitudes were 11.37, 11.97, and 11.38.  These magnitudes presumably date from around 1991.  Like all the brightnesses in this subsection, these refer to the entire nebula, with near-zero contribution from the central star.
	
	The {\it 2MASS} magnitudes were J=12.098$\pm$0.034, H=12.248$\pm$0.048, K=11.506$\pm$0.029 (Cutri et al. 2003).  The magnitudes are from May 2000.
	
	The {\it IRAS} fluxes were 0.65 Jy at 12$\mu$, 15.59 Jy at 25$\mu$, 8.05 Jy at 60$\mu$ and 3.39 Jy at 100$\mu$ (Parthasarathy \& Pottasch 1989). These observations were made from January 1983 to November 1983.  This spectral energy distribution has the obvious interpretation as thermal emission from dust in the shell, with an average temperature of 125 K (Parthasarathy \& Pottasch 1989). 
	
	The {\it Akari} satellite all-sky survey gives the 9$\mu$ flux to be 0.089$\pm$0.009 Jy and the 18$\mu$ flux to be 2.57$\pm$0.05 Jy (Ishihara et al. 2010).  These brightnesses come from a survey running from May 2006 until August 2007.
	
	The {\it WISE} all-sky survey provides magnitudes (on the Vega magnitude system) for the various WISE bands to be 11.144$\pm$0.0022, 10.373$\pm$0.019, 5.161$\pm$0.014, 0.902$\pm$0.01 for the W1 (3.4$\mu$), W2 (4.6$\mu$), W3 (12$\mu$), and W4 (22$\mu$) bands respectively.   The {\it WISE} survey ran from January 2010 to January 2011.
		
	The {\it ATCA} radio light curve, from 1991 to 2002, has the Stingray declining steadily in flux density from 63.6$\pm$1.8 mJy (Parthasarathy et al. 1993) to 48.8$\pm$1.5 mJy at 4800 MHz (Umana et al. 2008).  
			
\section{Images}

Roughly 50\% of classical PN have halos, faint roughly-circular shells far outside the the bright well-known nebulosity (Chu et al. 1987).  For example, NGC 6720 (the famous Ring Nebula) has the obvious shell (with outer dimensions 90"$\times$65") surrounded by a faint shell extending 162"$\times$147" filled with fairly uniform mottling and arcs, with this outer halo being bright in {\it WISE} images.  Corradi et al. (2003) and Chu et al. (1987) have catalogs with pictures and intensity profiles.  The typical surface brightness is a thousandth of that of the inner classical PN.  The estimated ages for these outer shells are many tens of thousands of years as based on expansion velocities that default to 20 km s$^{-1}$ when not otherwise known.  Corradi et al. (2003) call these faint shells structures outside the classic PN as `AGB halos', where the ordinary stellar wind of the star during its AGB phase has become ionized.  They claim that the outer edge of the AGB halo results from the last thermal pulse in the AGB star.

It is possible that the Stingray might have an AGB halo outside its small classical PN shell.  Such a shell might be too large to be visible in the {\it HST} images.  So we thought it worthwhile to search for any AGB halo around the Stingray.  We examined deep images from DECam, SHASSA, 2MASS, and {\it WISE}, as reported below.  We have also examined the various available images from the Digital Sky Survey and from {\it HST}, but our negative results for any shell structure or circular arcs centered on the Stingray do not provide useful constraints.  No images were taken with {\it GALEX} or {\it Swift}.

\subsection{DECam Images}

Our DECam images in u', g', r', and i' were examined for any nebulosity surrounding the Stingray.  In our images, the classical planetary nebula is entirely in the near-saturated inner core of the normal stellar image.  So we are only sensitive to any outlying outer shell.  Our examination was visual, mainly because there is no way to automate or quantify a shell search when we have no idea of the size or shape.  We have very long experience at searching for shells or light echoes around stars, and we know that the human eye/brain combination is very sensitive to detecting shells.  Indeed, for irregular shells, visual examination is greatly better at detecting any nebulosity by pushing down towards the background noise limit.  With this, we detect no shell or nebulosity in any of our DECam images.  The r' image would record any H$\alpha$ emission, while the g' image would record any [OIII] emission.  

\subsection{SHASSA Images}

The Southern H-alpha Sky Survey Atlas (SHASSA) is a robotic wide-angle CCD survey in the H$\alpha$ line covering declinations south of +15$\degr$ (Gaustad et al. 2001).  Images have pixels that are 47.64 arc-seconds on a side, with a sensitivity down to about 0.5 Rayleigh.  The Stingray appears prominently about 4$\degr$ from the center of their Field 38.  The point-spread-function has a FWHM of around 1 arc-minute, so all the known shell appears as a point source, and any detectable shell would have to be larger than several arc-minutes in radius.  The continuum-subtracted and smoothed image shows no shells, circular arcs centered near the Stingray, or any structure associated with the Stingray out to a radius of over 1 degree.

\subsection{2MASS Images}

There are also 2MASS observations of V839 Ara in J, H, and K bands taken in May of 2000.  The radius of emission in each of these three bands are less then 3".  No outer shell was seen in J, H, or K. 

\subsection{{\it WISE} Images}

The archival {\it WISE} images covers V839 Ara in four different bands centered on 3.3$\mu$$m$ (W1), 4.6$\mu$$m$ (W2), 12.1$\mu$$m$ (W3), and 22.2$\mu$$m$ (W4).  V839 Ara is positioned nearby a star (identified as TYC 8739-1088-1 with V=10.83) which is bright in both the W1 and W2 bands, however, it is noted that PNe are expected to show strong emissions in the far infrared. In the W1 band, the emission is centered on the location of V839 Ara with a radius of 11". Going further out into the infrared, the emission centered on V839 Ara extends out to a radius of 14" in W2, 34" in W4, and a finally a very large `halo' with a radius of 81" in W4. It should be noted that this `halo' seen out in 22$\mu$$m$ is consistent with PSF rings around sources bright in the W4 band, and is not astrophysical. This can be seen in Figure 3. 

\section{Light Curve Analysis for the Central Star}

From 1889 to 1980, the Stingray's central star exhibits a linear decline in its light curve, with substantial variability superposed.  The best fit line goes from 10.30 mag in 1889 to 10.76 mag in 1980, for a total drop of 0.46 mag.  This is the same as the apparent amplitude for the decade-long variations.  Over this time, with no substantial emission lines providing any significant fraction of the flux, the variability can only be caused by changes in the photosphere, either by variations in the effective temperature or radius.  (Changes in the dust column seem unlikely given that the stellar wind must be weak, while even large winds in the 1980's led to no change in the extinction.)  From {\it c.}1920 to 1980, the spectral type remained largely unchanged, with everyone reporting spectral types from B0 to B3, so the star's surface temperature was not changing by any large factor.  For the B-band flux to change by 0.46 mag, the radius would have to change by 20\%.  This is a large change.

How can the Stingray star have `anticipated' the 1980s ionization event?  Apparently, some prelude to the sudden event had a surface manifestation for at least a century in advance.  The anticipation cooling and/or shrinking of the central star could not have been going on for many centuries.  At its rate of fading by near half-a-magnitude per century, a millennium duration would imply that the central star was five magnitudes more luminous at the start of the anticipation.  Such would imply a luminosity higher than supergiants.  So this anticipation fading can only have been going on from one century to a few centuries.  Detailed calculations of the variations of the central star luminosity can have substantial changes in the century preceding thermal pulses (Blocker 1995; Sch{\"o}nberner 1983).  For a 0.553 M$_{\odot}$ star, before a thermal pulse at 45,000 K, the luminosity will drop by a factor of 2.2 in the preceding century (Sch{\"o}nberner 1983).  For a 0.836 M$_{\odot}$ star, the luminosity will drop by a factor of 6 in the century preceding a a thermal pulse at 200,000 K (Blocker 1995).  

Handler (2003) has defined a new class of variable stars, called the `ZZ Leporis' stars, which consist of the central stars of young PN with temperatures $<$50,000 K.  The physical mechanism for these brightness changes is not known, but it is likely some combination of stellar pulsations and fluctuations in a stellar wind.  Arkhipova et al. (2013) have already identified the central star of the Stingray as being in this ZZ Lep class.  However, although the variations are fast and aperiodic in both cases, the ZZ Lep stars have greatly different light curve properties than what we see for the Stingray.  In particular, ZZ Lep stars have time scales of 4-10 hours and amplitudes $<$0.03 mag (Handler 2003; Handler et al. 2013).  This is greatly different from the Stingray's behavior in 1889-1980, which has a time scale of around one decade and amplitude 0.5 mag.   From 1994-2009, our V-band light curve is not coming from the central star, while the central star may or may not have ZZ Lep behavior.

The fast decline for the light from the central star alone from 1980 (B=10.8) to 1988.5 (B=12.5) to 1996.3 (B=14.64) is stark and unprecedented.  This cannot be due to changing extinction, because the dimming, as measured by many methods, is essentially unchanged from 1980 to 2011 (Reindl et al. 2014).  So this fast fading star can only be due to some combination of the decrease of the stellar radius and a decrease in the temperature.  For the case of the Stingray, Reindl et al. (2014) show that the stellar temperature increased greatly while the star's radius decreased greatly from 1988 to 2006.  The radius and temperature changes run against each other.  So it takes a detailed calculation as to whether these observed changes translate into the observed magnitude changes.  If we go by the usual luminosity equation, we have $L\propto R^2 T_{eff}^4$, where $L$ is the luminosity, $R$ is the stellar radius, and $T_{eff}$ is the effective temperature of the photosphere.  We do not have direct measures of the radius, but the star's surface gravity, $g$, will scale as $g\propto R^{-2}$.  And we should not be using the luminosity, but rather the blackbody flux in the B-band, $F_{BB}$, at $\lambda$=4400\AA, with this being the usual Planck function of $\lambda$ and $T_{eff}$.  We can then get a B magnitude as $B=B_0-2.5\log (F_{BB}/g)$, for some zero magnitude $B_0$.  Reindl et al. (2014) list measured values for $\log$(g) and $T_{eff}$ for many years from 1988 to 2006.  Taking $B_0=43.55$ so that B=12.5 in 1988, we get B=13.1 mag in 1996, and B=14.9 in 2006.  We see that in this case for a shrinking star that is getting hotter, it is the shrinking that is the dominant effect, making for the star dropping in B-band brightness.  However, the effect predicted from the measured $\log$(g) and $T_{eff}$ is a much slower decline that is actually observed.  That is, from 1988 to 1996, the prediction is that the central star will dim by 0.6 mag, while the observed dimming is by 2.14 mag.  This is a large difference.  This points to the central star having additional light above the photosphere in 1988 that went away by 1996, or to changing systematic errors in measures of the surface gravity (perhaps associated with the changing stellar wind rate).  From this analysis, we take the basic cause for the fading of the central star to be due to the shrinking of the stellar radius, although detailed calculations do not predict the central star to be fading as much as is observed.

\section{The Fading Nebula}

The V-band light curve (essentially the nebula's brightness) from 1994 to 2015 has been steadily fading at the rate of 0.090 mag/year.  This is a highly significant result from three independent sources and has no prospect of being due to any artifact.  So we have the [OIII] emission line flux fading with a half-life of around 8 years.  From 1994 to 2015, the Stingray emission lines have faded by near 2.0 mag, a factor of near 6$\times$.  This is the nebula fading, not the central star.  

The fading of the nebula is also seen for wavelengths from the near infrared, the middle-infrared, and the radio:  (1) From around 1991 to 2000.4, the Stingray faded by 0.73, 0.28, and 0.13 mag for the J, H, and K bands respectively.  These correspond to half-lives of 9.3, 24, and 54 years respectively.  (2) We can use the {\it IRAS}, {\it Akari}, and {\it WISE} mid-IR fluxes to chart the changes in the 12$\mu$ flux.  For this, the {\it WISE} magnitude must be converted to Jansky, and the {\it Akari} flux must be interpolated to 12$\mu$.  With this, we have 12$\mu$ fluxes of 0.65 Jy in 1983.5, 0.36 Jy in 2007.0, and 0.226 Jy in 2010.6.  This shows a steady decline over 27.1 years with a half-life of 20.4 years.  (3)  Similarly, for 22$\mu$, we have fluxes of 8.96 Jy in 1983.5 and 3.20 Jy in 2010.6.  The extrapolation of the {\it Akari} 18$\mu$ flux to 22$\mu$ is just where the spectral energy distribution is turning over, so the uncertainty is too large for this to be useful.  This shows a decline in flux with a half-life of 18 years. (4) From 1991.3 to 2002.7, the ATCA radio flux at 4800 MHz (6 cm) declined from 63.6 mJy to 48.8 mJy.  This corresponds to a half-life of 30 years.

	We see a consistent picture for the light curves from 1994-2015, where the optical, near infrared, middle-infrared, and radio fluxes have all been smoothly fading with a characteristic time scale of a decade or so.  The nebula had a sudden turn-on sometime between 1979.49 and 1988, and the brightening of the nebula likely stopped in the early 1990's with the turn-off of the fast stellar wind associated with the 1980s ionization event.  The fading of the nebula started after the end of the ionizing event and its fast stellar wind, in the early 1990s.  At this time, the central star underwent fast fading (see Figure 1), so its illumination of the nebula started to decline fast.  It is easy to ascribe the fading of the nebular light to the sharp fall off in its illumination by the central star.  Even with the rise of surface temperature of the star, the fall in its surface area means that it is giving off much less ionizing radiation.  With the turnoff of the illuminating source, the nebula light should start fading at some sort of a re-ionization time scale.  There is precedent for this fading and this interpretation for Sakurai's Object (V4334 Sgr, a very-late-thermal-pulse born-again star).  Its emission line fluxes have also been seen to decline fairly rapidly (with a half-life just under two years) from 2001 to 2006, with the fading attributed to cooling and recombination in the shell after the heating and ionization ended (van Hoof et al. 2007).

	For the usual nebula conditions, the recombination time scale is around 80,000 years divided by the electron number density (in units of cm$^{-3}$).  (The nebula will have a range of densities, each recombining and fading on their own time scale, with continual-but-fading illumination from the central star, so the connection between the average electron density and the effective recombination times and the fading rate has substantial uncertainty and change over time.)  For the Stingray, Parthasarathy (2000) gives an electron density of 10,000 cm$^{-3}$, so we expect a recombination time scale of around 8 years.  The agreement between the half-life for the fading of the optical emission lines and the recombination time scale is an indication that the slow steady fading of the nebular light from 1994-2015 is caused by the ordinary recombination inside a nebula that had some sort of an ionization event between 1980 and 1994.
	
	The middle infrared emission is dominated by thermal light from warm dust in the nebula.  The dust grains are not likely to be changing in number or size, so the thermal flux can only be fading due to a cooling of the dust grains.  This is all consistent with a scenario where the dust was heated during the 1980s ionization event and has been cooling ever since.  The dust reaches thermal equilibrium on a fast time scale, so the fading of the middle infrared emission is a measure of the fading of the heating radiation field.  We suggest that the dust heating is dominated by the nebular emission lines, with this being {\it in situ} and the connection is made because the nebular lines are fading at the same rate that the dust is fading.
	
	The radio emission is free-free light, with model derived electron densities of 1.23-2.5 $\times$ 10$^4$ cm$^{-3}$ (Parthasarathy et al. 1993; Umana et al. 2008).  Like for the optical line emission, with a scenario of a short-duration ionization event in the 1980s, the radio flux should fade on some sort of a recombination time scale.

\section{Evolution Through the HR Diagram}

With the results from this paper, we can construct a detailed evolutionary path of the Stingray through the HR diagram.  For seven dates, we have calculated $L$ and $T_{eff}$ for the central star (see Table 3).  These can be directly plotted onto the HR diagram and compared to model predictions (see Figure 4).  The first point, labeled for the year 1001 AD, is meant to illustrate the approximate position where the star ejected the PN shell, just as it was leaving the AGB phase.  This is useful to show that the Stingray took only around one millennium to cross from the AGB to the thermal pulse region.  The two points for 1889 and 1980 have a modest uncertainty from the distance, so they can be moved up or down in the diagram, but they must be moved up or down together.  Similarly, the last four points can be moved up and down together by modest amounts as the assumed stellar mass changes from the adopted 0.55 $M_{\odot}$.  The end result is an observed evolutionary path from when the Stingray leaves the AGB until sometime during a thermal pulse.

The default or common idea is that the Stingray central star has evolved off the AGB with the usual nearly horizontal path across the top of the HR diagram (e.g., Parthasarathy et al. 1993), and is now undergoing a late thermal pulse (e.g., Reindl et al. 2014).  (See the next section for possible alternatives.)  Late thermal pulses are when a shell of helium, just outside the carbon-oxygen core, is ignited, with the energy from this burning making the star temporarily increase back to giant size, with the result that the star forms a loop in the HR diagram.  Detailed evolutionary paths through the HR diagram depend on the stellar mass, while the thermal pulses can occur anywhere along the path (e.g., Sch{\"o}nberner 1983; Blocker 1995; Blocker 2001; Sch{\"o}nberner 2008).  For illustrative purposes, one such late thermal pulse evolutionary track (from Sch{\"o}nberner 1983) has been superposed on the Stingray's observed evolutionary track in Figure 4.  This track was chosen because it is for a stellar mass that might be similar to that of the Stingray, because the thermal pulse is at a similar temperature as for the fast evolution of the Stingray, and because the track has conveniently labelled time tick marks to allow direct comparison.

The comparison between the observed and theoretical evolutionary paths in Figure 4 shows fundamental problems:  (1) The time scale for the evolution from a temperature of order 5000 K (when the PN shell was ejected) to a temperature of 50,000 K (in 1996) is observed to be around one millennium, while theory dictates a time scale more like 13 millennia to cross from roughly 5000 K to 50,000 K.  This discrepancy can be resolved if the star is greatly more massive than 0.55 $M_{\odot}$, but this then raises other difficult discrepancies.  (2) The Harvard plates show the Stingray was fading from 1889 to 1980, while the spectra from 1920 to 1979.49 show a nearly-unchanging spectral type, so we have a distinct vertical segment on the HR diagram lasting about a century.  From the theoretical paths, this could only match the initial turn in to a thermal pulse, like from the tick marks ``13" to ``13.1" in the theoretical path shown in Figure 4.  But if this match be made, then the further observed path to 1988-2006 is mystifying.  (3) The Harvard plates show a sharp drop in the central star's luminosity from 1980 to 1988, all with the temperature not changing greatly, so we must have another connected nearly vertical segment which lasts only 8 years.  This segment does not match any theoretical segment, in particular because its 8 year duration (0.008 millennia for comparison with the theoretical tick marks) is many orders-of-magnitude faster than any expected evolution, even for a massive star.  (4) The 1988-1996-2002-2006 segment shows significant and complex evolution.  In under two decades, the central star has moved in the HR diagram by more than theoretical paths allow for within two centuries at its fastest.  And the total change in luminosity from 1889 to 2006 (around 1.65 in log-units) is greatly larger than allowed in theoretical models.

\section{Fundamental Problems}

We can point to two major discrepancies between measures of fundamental parameters for the Stingray:  First, the mass of the star has contradictory evidence pointing to 0.55 M$_{\odot}$ (Parthasarathy et al. 1995), 0.59 M$_{\odot}$ (Bobrowsky 1994), 0.354$^{+0.14}_{-0.05}$ M$_{\odot}$  (Reindl et al. 2014), or 0.87 M$_{\odot}$ with an initial mass up to 6 M$_{\odot}$  (Reindl et al. 2014).  Second, the variations in the measured temperature and surface gravity for the central star {\it qualitatively} reproduce the sharp decline in brightness from 1980 to 2006, but {\it quantitatively}, the predicted decline from 1988 to 1996 is 0.6 mag, in stark contrast to the observed 2.14 mag decline (Section 2.6).

A fundamental problem posed by the Stingray is the nature of the 1980s ionization event.  We have seen no paper that addresses the question of what is really going on during this event, likely because answers are not known to the fundamental questions.  (1) Why should the ionization event start so suddenly within a time scale of a few years?  The evolution time scale for the increase in ionizing flux resulting from a simple transit across the top of the HR diagram is many millennia, while the evolution time scale for a thermal flash is centuries and longer.  (2) What is the physical mechanism that makes for the high luminosity of ionizing flux?  Presumably the pre-existing shell was ionized by ultraviolet or far-ultraviolet radiation, and such might come from the central star getting hotter.  But the surface temperature of the star only changed from around 30,000 K (from 1920-1979.49) to 38,000 K (in 1988), and we need a detailed calculation to see whether such a modest temperature change can make the nebula evolve from only weak Balmer emission lines to domination of the spectrum by very bright high-ionization emission lines.  In any case, the simple use of the central star to provide the ionizing flux does not work because it heated to 60,000 K in the year 2002, but the shell was already fading due to recombination.  (3)  Why is the ionization event concurrent with the fast and heavy stellar wind as seen from 1988 to the early 1990s?  The time coincidence between wind and ionization strongly implies a causal connection.  But did the wind make the ionizing radiation, or did the ionizing radiation drive the wind?  (4) Why did the ionization event not produce any temporary brightening in the B-band light curve?  A mechanism that suddenly produces a large luminosity of ultraviolet light will almost-certainly produce blue light, but there is no flare in the Harvard light curve with good resolution throughout the entire time period when the flare should occur.  (5) Why should the duration of the ionizing event be less than one decade and why should the turn-off time be only a few years?  All the evolution time scales are much longer than a decade.   

A higher fundamental problem of the Stingray is the overall nature of its evolution.  The general answer is that the star must be somehow traversing the HR diagram to go from the AGB phase to the upper left where it will soon enough end up moving down the white dwarf cooling track.  But the simple right-to-left traverse of the HR diagram (as depicted in textbooks) cannot explain the fast and complex evolution of the Stingray's central star (see Figure 4 and Table 3).  So either some additional mechanism is superposing some complex evolution, or the Stingray is not the post-AGB star that it seems.  Here, we will briefly discuss three possibilities:

	The first possibility to the nature of the Stingray's evolution is that it is an ordinary post-AGB star undergoing some sort of a thermal pulse, wherein a layer near the surface of the star suddenly ignites nuclear burning, and this influx of energy will puff up the star's outer envelope to giant portions.  Thermal pulses will make a star appear to go through loops in the HR diagram.  Perhaps the Stingray is caught just at the time of a thermal pulse, already having completed part of the loop?  Thermal pulses come in two types, called `late thermal pulses' (LTP) and `very late thermal pulses' (VLTP).  The later starts only after hydrogen burning has become extinct, and a critical feature of distinction is that the stellar surface is virtually free of hydrogen (Sch{\"o}nberner 2008).  The known VLTP examples are V605 Aql, Sakurai's object (V4334 Sgr), and perhaps FG Sge (Lawlor \& MacDonald 2003; Sch{\"o}nberner 2008).  But the Stingray certainly cannot be a VLTP star because it has only evolved off the AGB by roughly a millennium (when it ejected the PN shell), and because the central star has high abundance of hydrogen (as shown by the prominent Balmer absorption lines visible before 1980).  So for this solution, the Stingray would have to be an LTP star.  The LTP can make the star trace out a wide variety of loops through the HR diagram (Sch{\"o}nberner 1983; Blocker 1995; Blocker \& Sch{\"o}nberner 1997; Van Winckel 2003).  In an extreme case (see Figure 2 of Blocker \& Sch{\"o}nberner 1997), the star's track has nine reversals of direction in the HR diagram.  For all these published loops, none match the observed track of the Stingray, as detailed in the previous section.  None of the published LTP tracks can account for the Stingray, and this is the fundamental problem.  Part of this fundamental mismatch between observation and model is that the Stingray is moving in the HR diagram at much faster rates than is allowed by theory.  Gesicki et al. (2014) has found similar problems in accounting for the white dwarf mass distribution, and they have had to postulate an acceleration by a factor of 3 in the evolutionary time scales of Blocker (1995).  In his major review on post-AGB stars, Van Winckel (2003) says "the detailed description of the badly known external post-AGB mass loss on top of the mass loss from the nucleosynthetic consumption is crucial for the transition time estimates" for the star crossing the HR diagram in various phases.  It might yet be possible for some new model calculation to match the Stingray's evolution, but until then, the LTP idea has a fundamental problem.
	
		The second possibility is that the Stingray is a product of a common envelope ejection of the outer envelope of a giant star.  That is, in ordinary binary evolution involving a red giant star, a binary can get entangled inside a common envelope, the stars will spiral close together, and the main sequence star can eject the outer envelope of the red giant, leaving a bare stellar core.  The ejected envelope will expand, and the hot red giant core will ionize the shell.  Reindl et al. (2014) suggested this post-common-envelope (post-CE) state for the Stingray, with the evidence being that the apparent luminosity was consistent with the models of Hall et al. (2013).  Hall et al. make a variety of predictions that can be used to distinguish the post-CE systems.  (1) The luminosity of the post-CE stars will always be less than about 3.16$\times$10$^3$ L$_{\odot}$ while traversing across the HR diagram from right to left.  In sharp contrast, the Stingray had a luminosity of 5.4$\times$10$^3$ L$_{\odot}$ in 1889.  It is only after the central star has faded by an order-of-magnitude does the observed luminosity come into agreement with the post-CE tracks.  It is possible to overcome this argument by merely having the Stingray at a closer distance, with $<$1.2 kpc being adequate and within the uncertainty for the distance.  (2) The post-CE central star would be a close binary with an orbital period from a fraction of a day up to weeks.  We know of no observational tests for this prediction.  (3) The evolution through the post-CE phase is greatly slower than for an LTP.  Hall et al. calculate that the time scale from the ejection of the shell until the central star is 30,000 K is around 10,000 years, while the time scale for the small turn in the motion along the HR diagram (with the central star hotter than 30,000 K) is 0.1 to 10 million years.  This model prediction is greatly against the observed time scales of one millennium from ejection to 30,000 K and decades for the Stingray to turn around in the HR diagram.  The stark failure of this post-CE model prediction makes for a strong argument against the second possibility.
		
		The third possibility is that the Stingray is not a PN at all, but is rather a `PN mimic' (see Frew \& Parker 2010 for a review).  This has been casually suggested by Zijlstra (2015) in just one short sentence.  A PN mimic is a star with a surrounding ionized shell that does not come from the ejection of the outer envelope of an AGB star.  PN mimics include Wolf-Rayet stars, `young stellar objects', symbiotic stars, B[e] stars, Herbig-Haro objects, reflection nebulae, diffuse HII regions, old novae, and supernova remnants.  With such a wide diversity of mimics, it is difficult to absolutely reject all mimic classes for the Stingray.  Nevertheless, the observed properties of the Stingray have little in common with any of the classes of mimics.  In detail, Frew \& Parker point out how optical emission line ratios can discriminate between a typical PN and the various types of mimics.  For this, taking the line fluxes from Parthasarathy et al. (1993), for the particular lines as defined by Frew \& Parker, we have $\log(F_{H\alpha}/F_{[S II]})$=1.83, $\log(F_{H\alpha}/F_{[N II]})$=0.48, $\log(F_{[N II]6584\AA}/F_{H\alpha})$=-0.59, and $\log(F_{[O III]}/F_{H\beta})$=0.97.  With these, we can look in their Figures 4 and 5, seeing that the Stingray is right in the middle of the PN region, and is far away from the various mimic class regions.  That is, the Stingray's emission line spectrum is a classical PN spectrum, and greatly different from the mimics.  Further, many properties of the Stingray are characteristic for PNe and are not found in many classes of mimics; including the position of the central star on the HR diagram near B1 II, a classic bipolar point-symmetric shaped shell, the $\sim$10 km s$^{-1}$ expansion velocity of the shell, dust with a temperature of 125 K, thermal free-free radio emission, a distance of around 340 pc from the galactic plane, and the lack of any nearby ISM clouds or young stars.  With all this, we have strong confidence that the Stingray is not a PN mimic.
		
		After considering all proposals for the nature of the Stingray's evolution, we have eliminated all but one, with the remaining idea being the common idea that the Stingray is a post-AGB star traversing the upper HR diagram while undergoing some sort of a thermal pulse.  The fundamental problem is that there is no published model for a thermal pulse that come anywhere near the fast time scale for evolution or the large vertical motion in the HR diagram.

\section{Critical Observations}

We can propose four sets of observations that are currently ongoing or feasible in the near future.  First, it would be good to get optical and ultraviolet spectra of the shell and the central star, so as to measure the temperature while keeping up with the fast evolution of the system.  For this, a program with the {\it HST} in Cycle 22, with N. Reindl as the principal investigator, will get near- and far-ultraviolet spectroscopy with the {\it COS} instrument.  Second, further {\it HST} images in 2016 would provide a long enough time baseline so as to measure an accurate expansion age for the ordinary PN shell.  These same {\it HST} images might show the fast wind from the 1980s event.  This short duration wind has a velocity and total mass like that of the shell ejected by the recurrent nova T Pyx, for which we see the impact of the wind onto a prior more-massive shell (Schaefer, Pagnotta, \& Shara 2010).  So we might expect to see this impact lighting up the inside of the old PN shell with bright emission lines.  New {\it HST} images will also allow for resolving the central star and placing it onto the HR diagram, so we can see the evolution from 2006-2016.  For this second set of observations, we have just been awarded time during Cycle 23 of {\it HST}, with Z. Edwards as principal investigator, to use the {\it WFC3} to take all these needed images.  Third, the outer halo of the Stingray can be sought in the radio regime (c.f. Oettl et al. 2015).  Fourth, time series spectra exposed to show photospheric absorption features of the central star might show a sinusoidal radial velocity curve that will discover a close-in companion star (as needed to shape the bipolar nebula) as well as define the separation and mass of the close-in companion.  This observing task could also discover a close binary remnant of a post-CE system.

~

~

This research has made use of the {\it APASS} database, located at the {\it AAVSO} web site. Funding for {\it APASS} has been provided by the Robert Martin Ayers Sciences Fund.   We thank A. Henden for help in getting the {\it APASS} time series.  We thank H. Bond and M. Parthasarathy for comments on our manuscript.

{}

\begin{deluxetable}{lll}
\tabletypesize{\scriptsize}
\tablecaption{Stingray timeline
\label{tbl1}}
\tablewidth{0pt}
\tablehead{
\colhead{Year}   &
\colhead{Reference}   &
\colhead{Event}
}
\startdata

c. 1001	                 &	Reindl et al. (2014)	&	Ejection of visible PN shell (assuming 8.4 km/s velocity)	\\
1889	&	Section 2.1	&	B=10.1,  fading by 0.0051 mag/year for 1889-1981	\\
c. 1920	&	HD catalog	&	B0 spectrum, no emission lines	\\
c. 1950	&	Henize (1976)	&	H$\alpha$ emission line only, no O[III]	\\
1971	&	Parthasarathy et al. (1995)	&	B1 or B2, class I or II, only emission is very weak H$\beta$	\\
1979.49	&	Drilling \& Bergeron (1995)	&	O or B spectrum, no emission line	\\
1980	&	Section 2.1	&	B=10.8, starts fast fade at 0.20 mag/year	\\
1988.5	&	Section 2.1	&	B=12.5, central star is fading fast	\\
1988	&	Feibelman (1995)	&	Bright UV continuum with P Cygni profile	\\
1990, 1992	&	Parthsaranthy et al. (1993; 1995)	&	Very bright emission lines dominates optical spectrum	\\
1992-1994	&	Feibelman (1995)	&	Bright UV emission lines with no P Cygni profile	\\
1996.3	&	Section 2.5	&	B=14.64, V=14.96 for central star alone	\\
1994-2015	&	Sections 2.2, 2.3, \& 2.4	&	V-band (i.e., [OIII] lines) fades steadily by 0.081 mag/year	\\

\enddata

\end{deluxetable}

\begin{deluxetable}{llllll}
\tabletypesize{\scriptsize}
\tablecaption{Stingray light curve 1889-2014
\label{tbl2}}
\tablewidth{0pt}
\tablehead{
\colhead{Julian Date}   &
\colhead{Year}   &
\colhead{Band}   &
\colhead{Magnitude}   &
\colhead{Sigma}   &
\colhead{Source}
}
\startdata

2411191.6463	&	1889.520	&	B	&	10.1	&	0.10	&	HCO (B 3743)	\\
2411895.7797	&	1891.448	&	B	&	10.5	&	0.10	&	HCO (B 6263)	\\
2411896.7927	&	1891.450	&	B	&	10.1	&	0.10	&	HCO (B 6284)	\\
2412637.6167	&	1893.479	&	B	&	10.5	&	0.10	&	HCO (B 9644)	\\
2412637.6247	&	1893.479	&	B	&	10.3	&	0.10	&	HCO (B 9645)	\\
2412994.6657	&	1894.456	&	B	&	10.5	&	0.10	&	HCO (B 11416)	\\
2413055.6143	&	1894.623	&	B	&	10.4	&	0.10	&	HCO (B 11901)	\\
2413293.8362	&	1895.275	&	B	&	10.5	&	0.10	&	HCO (B 12968)	\\
\enddata

\end{deluxetable}

\begin{deluxetable}{lll}
\tabletypesize{\scriptsize}
\tablecaption{Stingray evolution
\label{tbl3}}
\tablewidth{0pt}
\tablehead{
\colhead{Year}   &
\colhead{Log($T_{eff}$)}   &
\colhead{Log($L/L_{\odot}$)}
}
\startdata

c. 1001	&	3.70	$\pm$	0.20\tablenotemark{a}	&	3.74	$\pm$	0.5\tablenotemark{a}	\\
1889	&	4.48	$\pm$	0.10\tablenotemark{b}	&	3.74	$\pm$	0.41\tablenotemark{e}	\\
1980	&	4.45	$\pm$	0.11\tablenotemark{c}	&	3.38	$\pm$	0.41\tablenotemark{e}	\\
1988	&	4.58	$\pm$	0.13\tablenotemark{d}	&	2.65	$\pm$	0.75\tablenotemark{d,f}	\\
1996	&	4.70	$\pm$	0.10\tablenotemark{d}	&	2.73	$\pm$	0.67\tablenotemark{d,f}	\\
2002	&	4.78	$\pm$	0.08\tablenotemark{d}	&	2.24	$\pm$	0.63\tablenotemark{d,f}	\\
2006	&	4.74	$\pm$	0.09\tablenotemark{d}	&	2.09	$\pm$	0.65\tablenotemark{d,f}	\\
\enddata
\tablenotetext{a}{The temperature and luminosity for the epoch of the ejection of the planetary nebula shell are simply meant to represent the approximate position of the central star when it leaves the AGB.  The important realization is that the shell was ejected about a thousand years before the current situation, and this is greatly at odds with theoretical models that require an order of magnitude longer time, unless the central star is much more massive than anyone expects.}
\tablenotetext{b}{The central star was typed as B0 around 1920.}
\tablenotetext{c}{The central star was typed as B1 in 1971.}
\tablenotetext{d}{Reindl et al. (2014).}
\tablenotetext{e}{The luminosity is based on the observed B magnitudes (see Table 1), corrected for extinction (with $E(B-V)=0.20\pm0.04$), corrected to V magnitudes (with the $B-V$ color appropriate for the observed spectral type), converted to absolute V magnitude (with a distance of $1.6\pm0.6$ kpc), corrected to the absolute bolometric magnitude (with a correction appropriate for the observed spectral type), and converted to luminosity in solar units (with a scaling from our Sun).}
\tablenotetext{f}{The luminosity is calculated as $L=(\sigma T^4)(4 \pi GM_{star}/g)$, where $g$ is the star's surface gravity as measured by Reindl et al. (2014), and where the mass of the central star is taken to be the reasonable middle value of 0.55 $M_{\odot}$ (as appropriate for comparison with the evolution tracks of Sch{\"o}nberner 1983).}
\end{deluxetable}

\begin{figure}
	\centering
	\makebox[\textwidth][c]{\includegraphics[width=1.\textwidth]{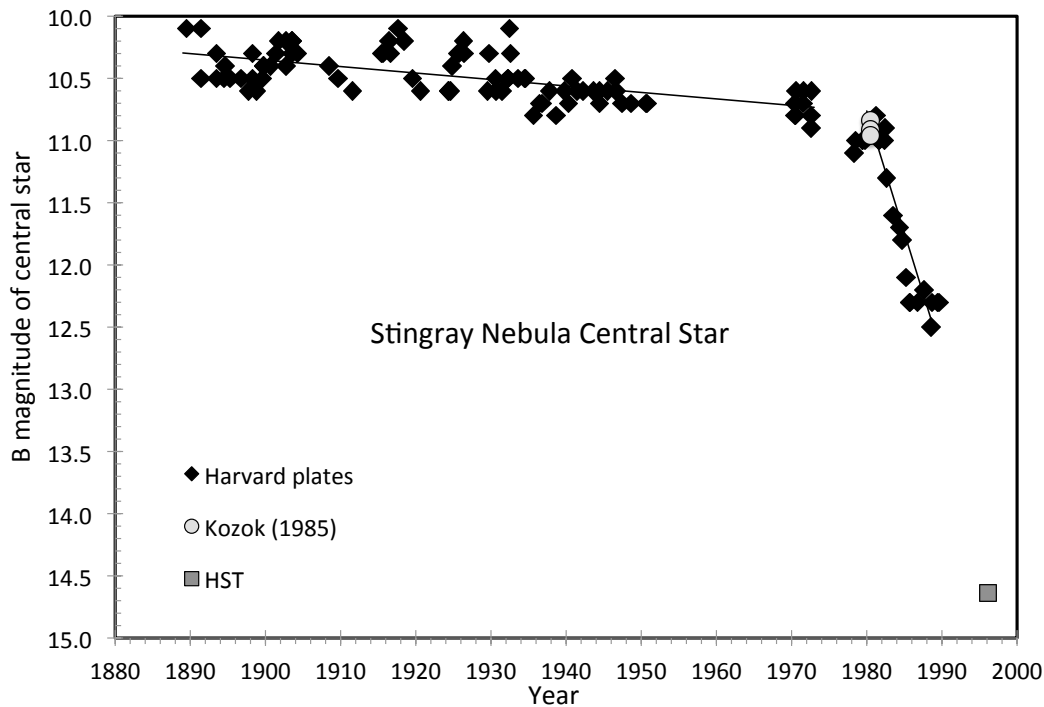}}
	\caption{B-band light curve of the central star of the Stingray Nebula.  The Harvard plates show a steady and significant decline from 1889 to 1980, followed by a sudden fast fading from 1980 to 1989.  The average rate from 1889-1980 is 0.0051 mag/year, as represented by the thin line.  So in some sense, the star knew to anticipate the upcoming ionization event in 1980.  Superposed on this linear decline are apparent variations on the time scale of a decade by up to half a magnitude in amplitude.  The B-band magnitudes from 1889-1980 are certainly of light from the photosphere of the central star, because the four spectra before 1980 show either no emission lines or very weak emission lines.  In 1980, the three B-magnitudes from Kozok (1985a) confirm the Harvard light curve.  After 1980, the B magnitude starts fading fast, with an apparent rate of 0.20 mag/year, as represented by the thin line.  This huge dimming is likely caused by the observed decrease in the size of the central star, although detailed calculations do not reproduce the speed of the decline.  If emission lines contributed significantly from 1980-1989, then this could only mean that the central star was fading even faster.  In 1996, {\it HST} resolved the central star from the surrounding nebulosity, while the B-band flux was taken from a spectrum without any emission contribution, so this magnitude is also of the central star alone.  The decline from 1980 to 1989 can be extrapolated to accurately reproduce the 1996 {\it HST} magnitude.  Importantly, there is no sign of any brightening in the B-band light curve around the time of 1980, when the ultraviolet flux suddenly turned on very brightly.}
\end{figure}

\begin{figure}
	\centering
	\makebox[\textwidth][c]{\includegraphics[width=1.2\textwidth]{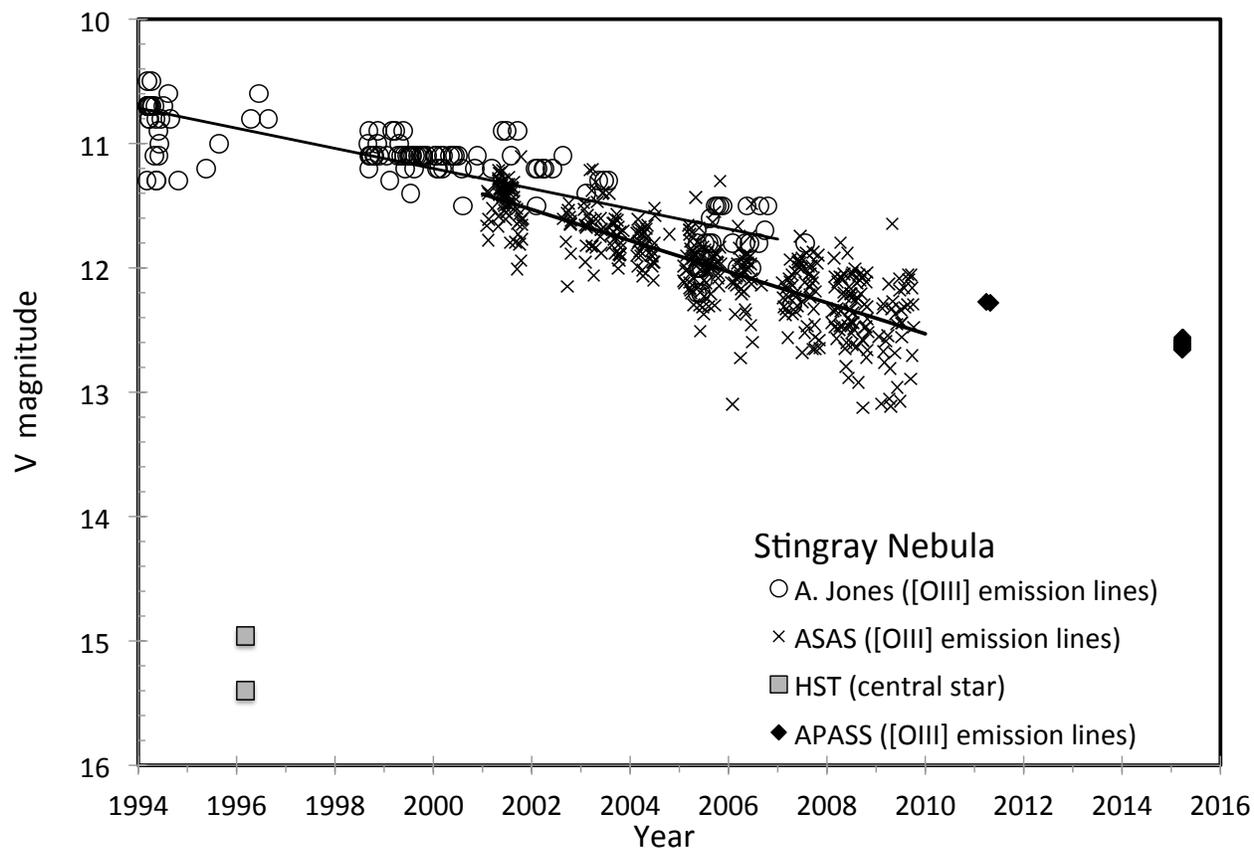}}
	\caption{V-band light curve of the Stingray from 1994-2015.  Albert Jones and {\it ASAS} have reported many V-band measures, and they are in reasonable agreement.  (Small differences are expected between observers due to small differences in their spectral sensitivity in looking at an emission line source as compared to the continuum of the normal comparison stars.)  They show a slow rate of decline (indicated by the thin lines), with rates of 0.081 mag/year and 0.125 mag/year.  Importantly, these V-band magnitudes are really just measuring the brightness of the two [OIII] emission lines at 5007\AA~ and 4959\AA, because contemporaneous spectra show that these lines provide 98\% of the detected light for the V-band.  Also importantly, contemporaneous {\it HST} narrow-band images centered on the [OIII] lines shows that the central star is very faint.  The V-band magnitude for the central star alone has been isolated with {\it HST} in 1996, with the star itself providing less than 2\% of the overall V-band flux.  }
\end{figure}

\begin{figure}
	\centering
	\makebox[\textwidth][c]{\includegraphics[width=1.\textwidth]{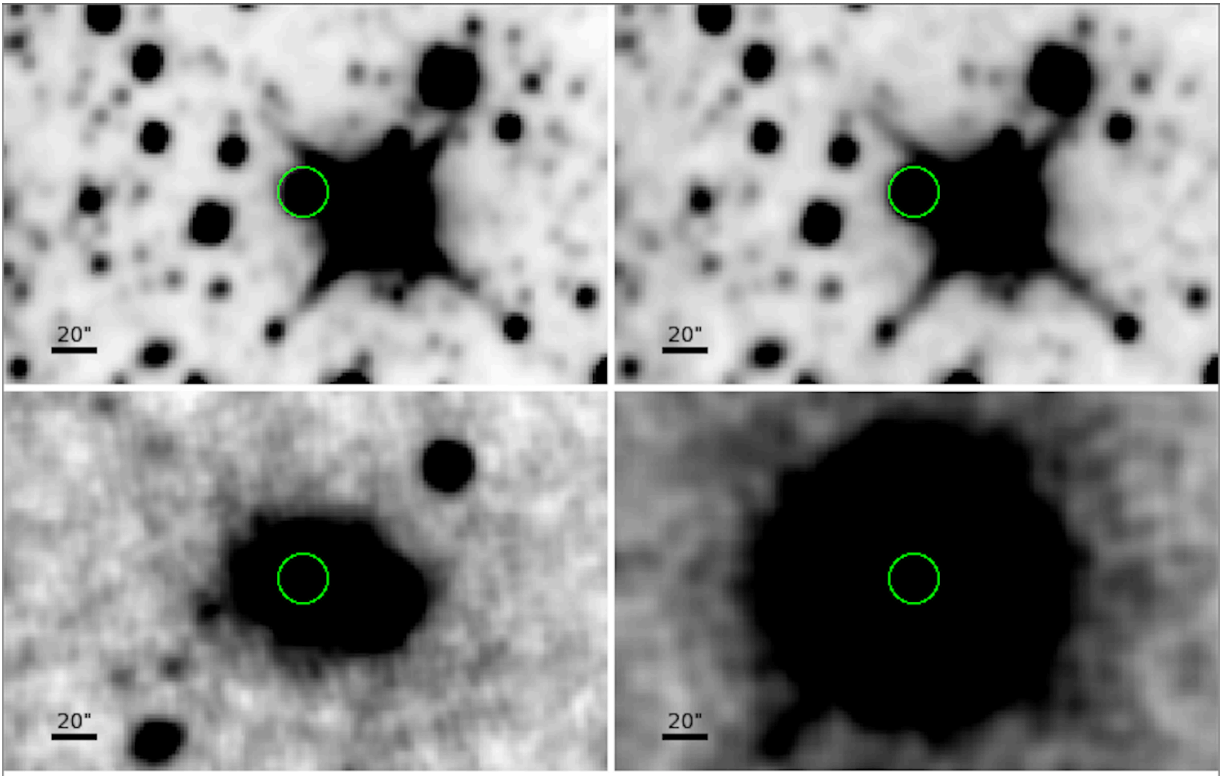}}
	\caption{WISE band 1 image (top left), WISE band 2 image (top right), WISE band 3 image (bottom left), WISE band 4 image (bottom right). V839 Ara is marked by the green circle on each image. The 81" `halo' seen in the WISE band 4 image is centered on V839 Ara and is consistent with PSF rings in W4 extending past $\sim$50". There is no physical shell associated with these infrared wings of the PSF. }
\end{figure}

\begin{figure}
	\centering
	\makebox[\textwidth][c]{\includegraphics[width=1.\textwidth]{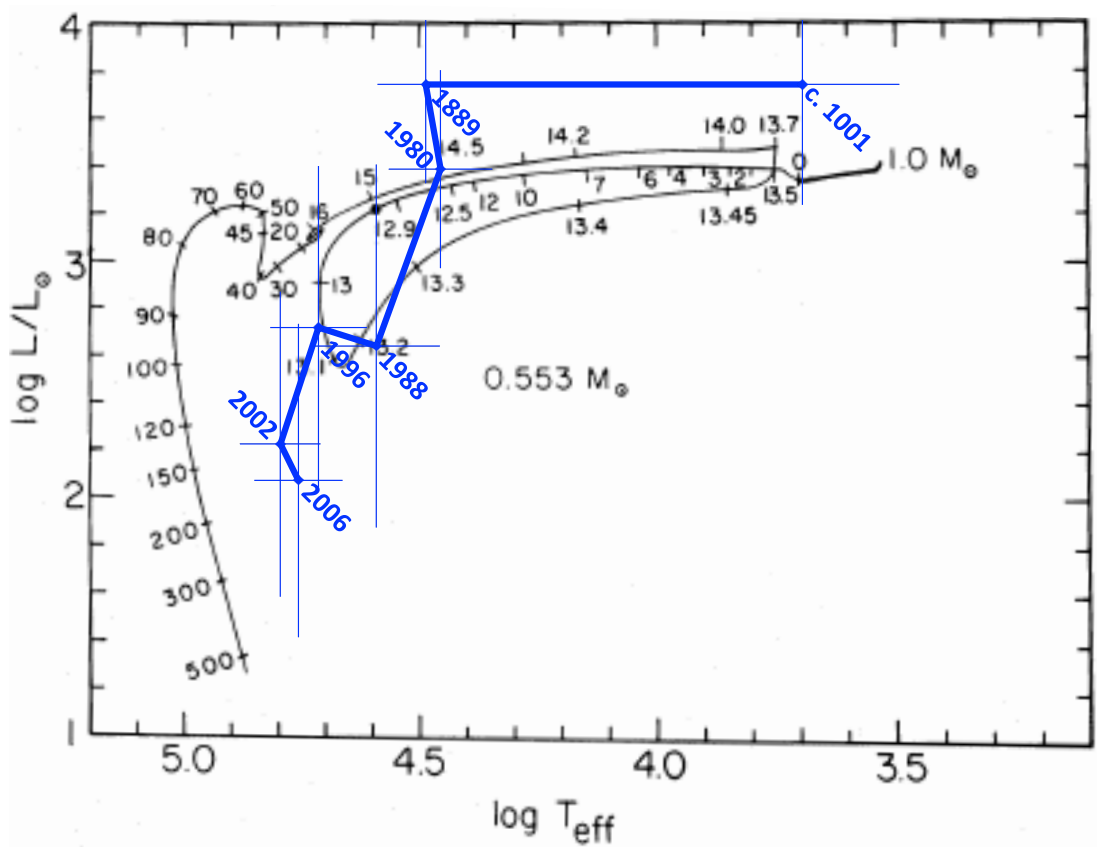}}
	\caption{The observed versus theoretical evolutionary paths for the Stingray.  The theoretical evolutionary path for a 0.553 $M_{\odot}$ star is shown as a black curve with tick marks showing the time since the start of the post-AGB phase in units of thousands of years, as copied from Figure 4 of Sch{\"o}nberner (1983).  The important points from this are that is takes around 13,000 years from the time of the shell ejection until the star has heated up to 50,000 K, that the time scale for the pulses is from centuries to millennia, and that the range of luminosity is just 0.8 in the logarithm.  The observed evolutionary path for the Stingray (the thick blue line with points labeled by the year) is taken from Table 3.  The placement of the point for 1001 AD is an approximation of its position when the shell was ejected, with the point being that the observed time scale from shell ejection to the thermal pulse is one order-of-magnitude smaller than theory allows (at least for a non-massive star).  The points for 1889 and 1980 can move up-or-down somewhat due to uncertainty in the distance to the Stingray, but the two points move together, so we are left with a vertical segment in the real evolution that does not readily match any model prediction.  The last four points (1988-2006) show a path that does not match any model predictions.}
\end{figure}

\end{document}